\documentclass[a4paper]{jpconf}
\usepackage{graphicx}
\usepackage{amssymb}

\begin{document}

\title{Dynamical decoupling of superconducting qubits}
\author{Jian Li}\ead{jian.li@aalto.fi}
\author{G. S. Paraoanu}
\address{Low Temperature Laboratory, Aalto University, PO Box 15100, FI-00076 AALTO,
Finland}

\begin{abstract}
We show that two superconducting qubits interacting {\it via} a fixed transversal coupling can be decoupled by appropriately-designed microwave field excitations applied to each qubit. This technique is useful for removing the effects of spurious interactions in a quantum processor. We also simulate the case of a qubit coupled to a two-level system (TLS) present in the insulating layer of the Josephson junction of the qubit. Finally, we discuss the qubit-TLS problem in the context of dispersive measurements, where the qubit is coupled to a resonator.
\end{abstract}

%\maketitle

%%%%%%%%%%%%%%%%%%%%%%%%%%%%%%%%%%%%%%%%%%%%%%%%%%%%%%%%%%%%%%%%%%%

\section{Introduction}

In the past years, fixed transversal couplings between two superconducting qubits have been extensively studied theoretically \cite{Rigetti, Ashhab, Li08} and experimentally in systems comprising phase qubits \cite{Berkeley03}, charge qubits \cite{Yu03}, flux qubits \cite{Groot10}, and in circuit QED systems \cite{Majer07}. Much of the motivation of these studies comes from the need of developing reliable techniques for modulating the coupling between qubits. This is required in order to produce in a controllable way CNOT quantum gates \cite{para} - the basic building blocks of quantum algorithms.

\section{Two qubits with fixed coupling}

In this paper we consider a system of two transversely coupled qubits under microwave driving.  For simplicity we take $\hbar = 1$. The Hamiltonian \cite{Groot10} can be written as
\begin{eqnarray}
H &=& -\frac{1}{2}\left( \Delta_1\sigma_1^z + \Delta_2\sigma_2^z \right) + J\sigma_1^x\sigma_2^x \nonumber \\
&& + \Omega_1\cos(\omega_d t + \varphi_1)\sigma_1^x + \Omega_2\cos(\omega_d t + \varphi_2)\sigma_2^x ,
\label{eq_transverse_ham}
\end{eqnarray}
where $\Delta_j$ is the energy splitting (Larmor frequency) of qubit-$j$, $J$ is the inter-qubit coupling strength, $\Omega_j$ and $\varphi_j$ indicate the amplitude (Rabi frequency) and the relative phase of the driving fields at the frequency $\omega_d$ for qubit-$j$, respectively, and $\sigma_j^{x,y,z}$ are the qubit-$j$ Pauli matrices in the undriven energy eigenbasis.

\subsection{Effective Hamiltonian}

\begin{figure}[htb]
\begin{center}
\includegraphics[width=9cm]{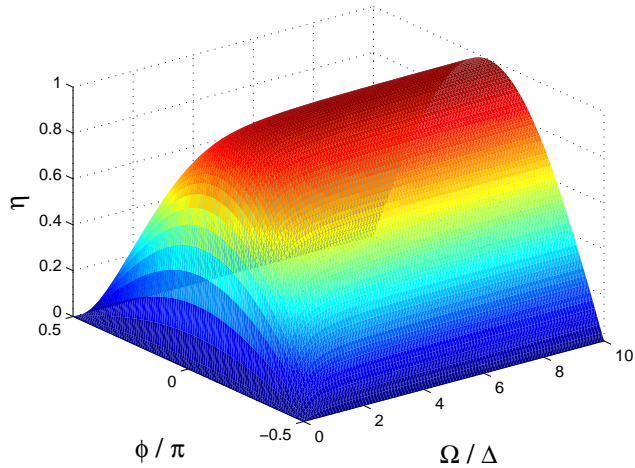}
\end{center}
\caption{(Color online) The dimensionless coupling strength $\eta$ as a function of $\Omega$ and $\phi$. } \label{fig_switchable_coupling}
\end{figure}

In the simple case when the driving frequency $\omega_d = (\Delta_1 + \Delta_2) / 2$, and the Rabi frequencies $\Omega_1 = \Omega_2 \equiv \Omega$, Eq. (\ref{eq_transverse_ham}) can be rewritten, following the same procedures as those in Sec. III of \cite{Li08}, as
\begin{eqnarray}
H_{\rm eff} = \frac{J}{4}\eta \left[ \sigma_{(1)}^x\sigma_{(2)}^x + \sigma_{(1)}^y\sigma_{(2)}^y + 2\sigma_{(1)}^z\sigma_{(2)}^z \right] ,
\label{eq_eff_ham}
\end{eqnarray}
where $\sigma_{(j)}^{x,y,z}$ are the Pauli matrices of qubit-$j$ in the driven energy eigenbasis, see \cite{Li08}. We obtain the dimensionless qubit-qubit coupling strength $\eta$ as
\begin{equation}
\eta = \frac{\Omega^2}{\Omega^2 + \Delta^2/4}\cos\phi ,
\end{equation}
where $\Delta = \Delta_1 - \Delta_2$ is the energy difference between the
two qubits, and $\phi = \varphi_1 - \varphi_2$ indicates the phase difference between
the two driving fields. It can be tuned by either $\Omega$ or $\phi$ independently, as
shown in Fig. \ref{fig_switchable_coupling}.
%We therefore have the flexibility to implement single-qubit gates by tuning $\Omega$ and fixing $\phi$ at $\pm \pi /2$.

The time evolution operator generated by $H_{\rm eff}$ is
\begin{equation}
{\cal U}_{\mathrm{eff}}(t) = \frac{1}{2} \left[
    \begin{array}{cccc}
        2 \exp(-i J\eta t / 2) & 0 & 0 & 0 \\
        0 & 1 + \exp(i J\eta t) & 1 - \exp(i J\eta t) & 0 \\
        0 & 1 - \exp(i J\eta t) & 1 + \exp(i J\eta t) & 0 \\
        0 & 0 & 0 & 2 \exp(-i J\eta t / 2)
        \end{array}
    \right] . \label{eq_time_evo_oper}
\end{equation}

%%%%%%%%%%%%%%%%%%%%%%%%%%%%%%%%%%%%%%%%%%%%%%%%%%%%%%%%%%%%%%%%%%%

% \section{Entangling property of the system}
% \label{RWA}

\subsection{Entanglement}

The entangling properties of a system of two qubits can be characterized by calculating an entanglement measure known as {\em concurrence} \cite{Wootters}, which is defined as
\begin{equation}
{\cal C}(\psi) = |\langle\psi|\sigma_1^y\otimes\sigma_2^y|\psi^*\rangle|  \label{eq_conc_def1}
\end{equation}
for a pure two-qubit state $|\psi\rangle$. Here $|\psi^*\rangle$ is the complex conjugate of $|\psi\rangle$. For a general two-qubit state $|\psi\rangle = c_{00}|00\rangle + c_{01}|01\rangle + c_{10}|10\rangle + c_{11}|11\rangle$, the concurrence is
\begin{equation}
{\cal C}(\psi) = 2|c_{00}c_{11}-c_{01}c_{10}| \leq 1 , \label{eq_conc_def2}
\end{equation}
where $|m n\rangle \equiv |m\rangle_1\otimes|n\rangle_2$, and $|0\rangle_j$ ($|1\rangle_j$) represents the ground (excited) state of qubit-$j$.

\begin{figure}[htb]
%\begin{center}
\includegraphics[width=8cm]{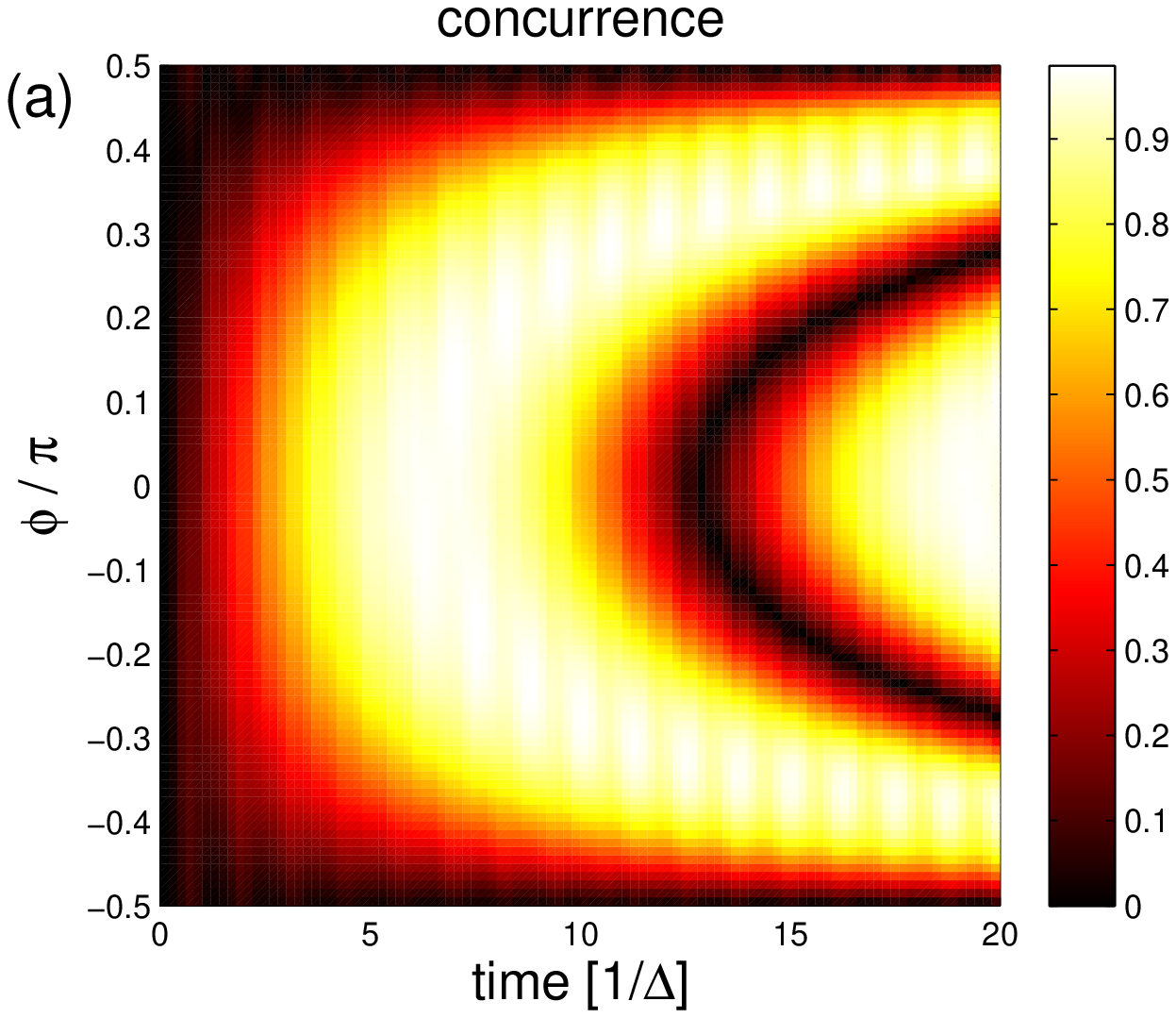} \includegraphics[width=8cm]{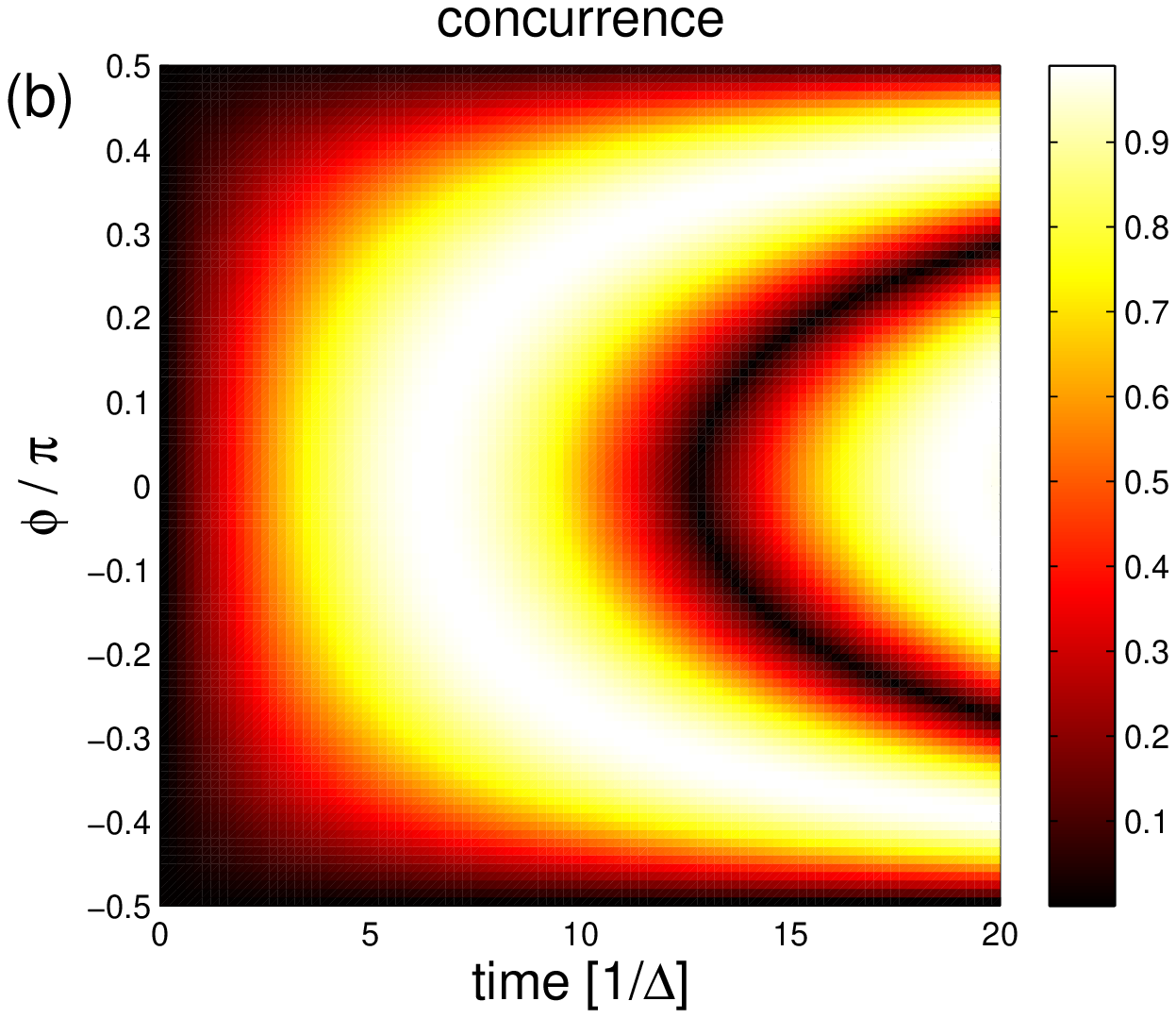}
%\end{center}
\caption{(Color online) The phase-dependent concurrence evaluated (a) by using the original Hamiltonian Eq. (\ref{eq_transverse_ham}) and (b) by using the analytical expression Eq. (\ref{eq_conc_eff}). The parameters used in this plot are $\Omega = 5\Delta$ and $J = \Delta/2$.}
\label{fig_concurrence}
\end{figure}

If initially the two qubits are in their ground states $|00\rangle$, by using the time evolution operator Eq. (\ref{eq_time_evo_oper}) and within the approximation that the Rabi frequency is much larger than the energy difference between the two qubits, $\Omega \gg \Delta$, the time-dependent concurrence Eq. (\ref{eq_conc_def1}) can be put in the following approximate form \cite{Li08}:

%\begin{equation}
%{\cal C}(t) = \frac{1}{4}\left| \left( e^{-i J\eta t} - e^{2i J\eta t} \right)\cos(\theta_1 - \theta_2) + e^{2i J\eta t} -1 + \left( 1 - e^{-i J\eta t} \right) \cos(\theta_1 + \theta_2) \right| ,
%\end{equation}
%where $\theta_j = \arctan(\Omega / \delta_j)$, with the detuning
%$\delta_j = \omega_d - \Delta_j = \pm \Delta / 2$.
%If the Rabi frequency is much larger than the energy difference between the two qubits, $\Omega \gg \Delta$, the concurrence is simplified to

\begin{equation}
{\cal C}(t) \approx \left| e^{i J\eta t} -1 \right| / 2 =  \left|\sin\left[
\frac{2J\Omega^{2}\cos\phi}{4\Omega^{2} + \Delta^{2}}t\right]\right|,
\label{eq_conc_eff}
\end{equation}
as shown in Fig. \ref{fig_concurrence} (b) below, which is a good approximation to the concurrence numerically calculated by using the original Hamiltonian Eq. (\ref{eq_transverse_ham}), as shown in Fig. \ref{fig_concurrence} (a).

%%%%%%%%%%%%%%%%%%%%%%%%%%%%%%%%%%%%%%%%%%%%%%%%%%%%%%%%%%%%%%%%%%%%

\section{Qubit-TLS systems}
% \label{qubit_tls}

A lot of experimental progress has been made recently on phase qubits following the realization that the dielectric insulator forming the Josephson junction contains two-level system (TLS) defects \cite{phase, junctionresonators}. These defects have been shown to have decoherence times comparable to that of the qubit, thus they can be addressed coherently ({\it e.g.}  by tuning the qubit on- and off- resonance with them).
The form of the interaction Hamiltonian between the qubit and the TLS is of the type
$\sigma^x\sigma^x$ in the case of phase qubits \cite{junctionresonators, ustinov}. The same type
of coupling is obtained in the case of charge-based qubits from TLSs located on the island
and in the case of flux qubits from pinning centers in the superconductors used for fabricating the qubits.

The interactions between a qubit and a TLS becomes relevant only when $\Delta \equiv |\omega_{\rm qb} - \omega_{\rm TLS}|\lesssim \kappa$, where $\kappa$ denotes the coupling strength between them, $\omega_{\rm qb}$ and $\omega_{\rm TLS}$ are transition frequencies of the qubit and the TLS, respectively. By assuming that for each single qubit there is only one such TLS near it, the Hamiltonian for this qubit-TLS system is written as
\begin{equation}
H_{\rm qb-TLS} = -\frac{\omega_{\rm qb}}{2}\sigma^z - \frac{\omega_{\rm TLS}}{2}\tau^z + \kappa\sigma^x\tau^x , \label{eq_ham_qubit_tls_0}
\end{equation}
with the TLS Pauli matrices $\tau^{x,y,z}$. To coherently control the qubit, we apply a transverse microwave field to it. Then the total Hamiltonian reads
\begin{equation}
H(t) = H_\mathrm{qb-TLS} + \Omega\cos(\omega_dt+\phi)\sigma^x . \label{eq_ham_qubit_tls}
\end{equation}

\begin{figure}[htb]
\begin{center}
\includegraphics[width=9cm]{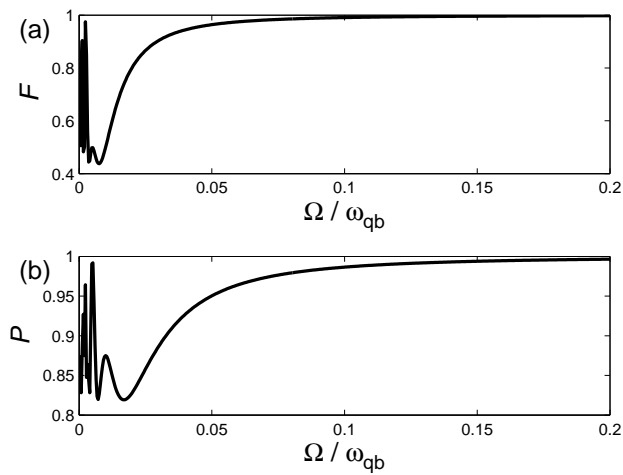}
\end{center}
\caption{(a) The fidelity ${\cal F}$ of the $\pi$-pulse as a function of driving amplitude $\Omega$. (b) The gate purity ${\cal P}$ as a function of the same variable. The qubit-TLS coupling is taken $\kappa = 5 \times 10^{-3} \omega_{\rm qb}$. }
\label{fig_fidelity_and_purity}
\end{figure}

In Fig. \ref{fig_fidelity_and_purity}(a) we have plotted the fidelity ${\cal F}$ \cite{Fidelity} of a $\pi$-rotation around the $X$-axis (see also Sec. V of \cite{Li08}), by simply taking $\omega_d = \omega_{\rm qb} = \omega_{\rm TLS}$, $\phi = 0$, and $\Omega$ to be time independent (rectangular pulse). The fidelity of a unitary transformation $U$ applied on the qubit between the initial pure state $|\psi_{\rm in}\rangle$ and the target state $\rho_{\rm out}$
is defined as ${\cal F}(U) = \overline{\langle \psi_{\rm in}|U^{\dagger}\rho_{\rm out}U|\psi_{\rm in}\rangle}$. Since we are not interested in the evolution of the TLS, when calculating the fidelity the output state $\rho_{\rm out}$ was obtained by tracing out the TLS degrees of freedom. Due to the qubit-TLS coupling, $\rho_{\rm out}$ is always a mixed state even if the input states is pure and unentangled. In other words, the entanglement between the qubit and the TLS results in decoherence for the qubit. As we can see from Fig. \ref{fig_fidelity_and_purity}(a) and expected on physical reasons, when the driving amplitude is much larger than the qubit-TLS coupling, the fidelity loss due to the TLS is negligible.

Since the state $\rho_{\rm out}$ is not pure, the gate purity ${\cal P} \equiv \overline{{\rm Tr}(\rho_{\rm out}^2)}$ \cite{Fidelity} should also be considered. In Fig. \ref{fig_fidelity_and_purity}(b) we show the numerical results of ${\cal P}$ for $\kappa / \omega_{\rm qb} = 5\times 10^{-3}$ \cite{Simmonds}. Again, as expected, for relatively large values of the
the driving amplitude compared to the coupling  $\kappa$, we find that the state becomes almost pure.

%For a single input state, $1/2\leq {\rm Tr}(\rho_{\rm out}^2) < 1$, averaged overall possible
%unentangled input states, the gate purity ${\cal P}$ should drop to $\sim 0.75$ for
%the pulse duration $T_2 \leq t \ll T_1$, where $T_1$ and $T_2$ are the relaxation
%time and the decoherence time of the qubit, respectively. In Fig. \ref{fig_fidelity_and_purity}(b) we show the numerical results of ${\cal P}$ for $\kappa / \omega_{\rm qb} = 5\times 10^{-3}$ \cite{Simmonds}, and the corresponding decoherence time $T_2 \approx 980 / \omega_{\rm qb}$ is about 15 ns when $\omega_{\rm qb} = 2\pi \times 10$ GHz.

\begin{figure}[htb]
\begin{center}
\includegraphics[width=8cm]{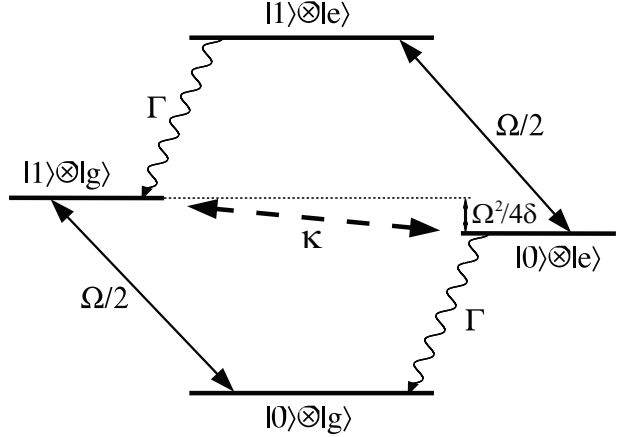}
\end{center}
\caption{Energy level configuration for the qubit-TLS system. $|0\rangle$ ($|g\rangle$) and $|1\rangle$ ($|e\rangle$) denote the ground and the excited states of the qubit (TLS). The solid arrows indicate qubit transitions with a rate $\sim g\sqrt{n}$ due to the coherent driving. The spontaneous decays of the TLS is indicated by the wiggly arrows, and the decay rate $\Gamma$ is assumed to be negligible compared with $\Omega$. The dashed double arrow denotes transitions due to the qubit-TLS coupling $\kappa$. }
\label{fig_qubit_tls_levels}
\end{figure}

\subsection{Qubit-TLS under dispersive measurement using a resonator}

For the qubit dispersively coupled to the resonator, it is possible to decouple the qubit and the TLS by driving the resonator. We take the Jaynes-Cummings form for the system Hamiltonian
\begin{eqnarray}
H = -\frac{\omega_{\rm qb}}{2}\sigma^z -  \frac{\omega_{\rm TLS}}{2}\tau^z + \omega_{\rm r} a^\dag a + g\left( \sigma^+ a + \sigma^- a^\dag \right) + \kappa (\sigma^+ \tau^- + \sigma^-\tau^+) , \label{eq_ham_qb_tls_res}
\end{eqnarray}
with $\sigma^\pm$ and $\tau^\pm$ the raising/lowering operators for the qubit and the TLS respectively,
%= (\sigma^x \pm i\sigma^y) / 2$, $\tau^\pm = (\tau^x \pm i\tau^y) / 2$,
$\omega_{\rm r}$ is the resonance frequency of the resonator, and $g$ is the  coupling strength between the qubit and the cavity mode.

In the dispersive regime the Rabi frequency $\Omega \equiv 2g\sqrt{n}$ ($n = \langle a^\dag a\rangle\gg 1$ indicates the number of photons) is much smaller than the detuning $\delta = \omega_{\rm r} - \omega_{\rm qb}$. To eliminate the qubit-photon coupling to leading order, we transform the Hamiltonian by using the Schrieffer-Wolff transformation operator $A = g(\sigma^+ a - \sigma^- a^\dag)/\delta$. Expanded to second order in $g/\delta$, the Hamiltonian is approximately
\begin{eqnarray}
 H' &=& e^{-A}H e^{A}
\approx H + [H,A] + \frac{1}{2}[[H,A],A] = \nonumber \\
&=&
-\left[ \frac{\omega_{\rm qb}}{2} - \frac{g^2}{\delta}(a^\dag a + 1/2) -
\frac{kg}{\delta}(\tau^+ a + \tau^-a^+)\right] \sigma^z - \frac{\omega_{\rm TLS}}{2}\tau^z + \nonumber  \\ & &  +\omega_{\rm r} a^\dag a + \kappa (\sigma^+ \tau^- + \sigma^- \tau^+) . \label{eq_sw_transform}
\end{eqnarray}
We now assume for the simplicity of the argument that the resonator is in a photon number state $|n\rangle$; then the term $\tau^+ a + \tau^-a^+$ (which can be interpreted as a qubit-mediated exchange of quanta between the resonator and the TLS) can be neglected and, up to a constant energy shift $n\omega_{\rm r}$,  we obtain
\begin{equation}
 H'\approx
-\left[ \frac{\omega_{\rm qb}}{2} - \frac{g^2}{\delta}(n + 1/2)
\right] \sigma^z - \frac{\omega_{\rm TLS}}{2}\tau^z + \kappa (\sigma^+ \tau^- + \sigma^- \tau^+) . \label{eq_sw_transform}
\end{equation}
 From this expression one sees that the qubit transition frequency is ac-Stark shifted by the quantity $\sim \Omega^2 / 4\delta = g^2 n/\delta$ due to the presence of  $n$ photons in the resonator. When $\Omega^2 / 4\delta\gg \kappa$, the transitions between the states $|0\rangle \otimes |e\rangle$ and $|1\rangle \otimes |g\rangle$, as illustrated in Fig. \ref{fig_qubit_tls_levels}, are suppressed. Therefore, in order to decouple the qubit and TLS, the driving field must satisfy $\kappa \ll \Omega \ll \delta$.

\section{Conclusions}

We have shown that by an appropriate choice of the amplitudes and phases of the microwave signals applied to a system of two qubits, the coupling between them can be modulated. In the case of a spurious coupling between a qubit and a TLS residing for example in the insulating layer of  the junction, this technique can be used for eliminating the decohering effect of the defect.

\section*{Acknowledgements}

This work was supported from NGSMP and Academy of Finland (projects 135135 and 141559).

%%%%%%%%%%%%%%%%%%%%%%%%%%%%%%%%%%%%%%%%%%%%%%%%%%%%%%%%%%%%%%%%%%%%

\end{document}